\documentclass{PoS}

\title{Correlations from generalized thermodynamic uncertainty relations}

\ShortTitle{Correlations from generalized thermodynamics}

\author{\speaker{Grzegorz Wilk}\\
        National Centre for Nuclear Research,
        Department of Fundamental Research, \\ul. Ho\.za 69, 00-681
        Warsaw, Poland\\
        E-mail: \email{wilk@fuw.edu.pl}}

\author{Zbigniew W\l odarczyk\\
        Institute of Physics, Jan Kochanowski University,\\
        \'Swi\c{e}tokrzyska 15, 25-405 Kielce, Poland\\
        E-mail: \email{zbigniew.wlodarczyk@ujk.edu.pl}}

\abstract{In order to account for possible nonstatistical
fluctuations in a hadronizing system while using a statistical
approach, one has to resort to its nonextensive version. The new
parameter $q$ is introduced to be directly connected to the
variance of an observable (with $q = 1$ one returns to usual
statistics). We demonstrate how this approach allows composing
fluctuations of different observables and show that it leads to a
specific sum rules proposing this to be verified experimentally.
We discuss the ensembles in which all relevant quantities are
allowed to fluctuate. By introducing correlations between the
observables, the relations connecting these variables are shown to
generalize the so-called Lindhard thermodynamic uncertainty
relations. This is illustrated using multihadron production data.
We show that fluctuations from different components of collision
phase space are correlated and that the strength of these
correlations depend on the relevant $q$-parameter.}

\FullConference{The Seventh Workshop on Particle Correlations and Femtoscopy\\
         September 20 - 24 2011\\
         University of Tokyo, Japan}

\begin{document}

\section{Introduction}

Nowadays it is a standard procedure to use a statistical approach
to model high energy multiparticle production processes
\cite{GGS}. However, it has also been realized that data on many
single particle distributions demonstrate a power-like behavior,
rather than the expected simple exponential one. In addition,
multiparticle distributions are broader than naively expected. The
proposition put forward some time ago is that, rather than
invalidating a simple statistical approach, these observations
call for its modification towards inclusion of a possible
intrinsic, nonstatistical fluctuations, usually identified with
fluctuations of the parameter $T$ identified with the
"temperature" of the hadronizing fireball (cf. reviews \cite{WW}).
The introduced parameter $q$, known as the {\it nonextensivity
parameter}, is shown to be directly connected to the variance of
$T$,
\begin{equation}
q = 1 + \omega_T^2\quad {\rm with}\quad \omega^2_T =
\frac{Var(T)}{<T>^2}, \label{eq:q}
\end{equation}
and one obtains the so called Tsallis distribution
\begin{equation}
h_q(E) = \frac{2-q}{T}\exp_q \left(-\frac{E}{T}\right) =
\frac{2-q}{T}\left[1 - (1-q)\frac{E}{T}\right]^{\frac{1}{1-q}} \,
\stackrel{q \rightarrow 1}{\Longrightarrow}\, \frac{1}{T}\exp
\left(-\frac{E}{T}\right). \label{eq:Tsallis}
\end{equation}
It reduces to the usual Boltzmann-Gibbs form for $q = 1$ (in which
case one recovers the usual statistical model)\footnote{This idea
has originated in \cite{WWq} and has been further developed in
\cite{BJ}. It forms a basis for so-called {\it superstatistics}
\cite{SuperS}. The problems connected with the notion of
temperature in such cases have been addressed in a recent book
\cite{B}. Cf. \cite{WW} for further references concerning
applications of this approach. Recent examples of spectacular
power-like behavior has been reported by PHENIX \cite{PHENIX} and
CMS \cite{CMS} recently.}. Such an approach is also known as
Tsallis statistics \cite{Tsallis}.

\section{Results}

\begin{figure}[h]
\begin{center}
\includegraphics [width=7.6cm]{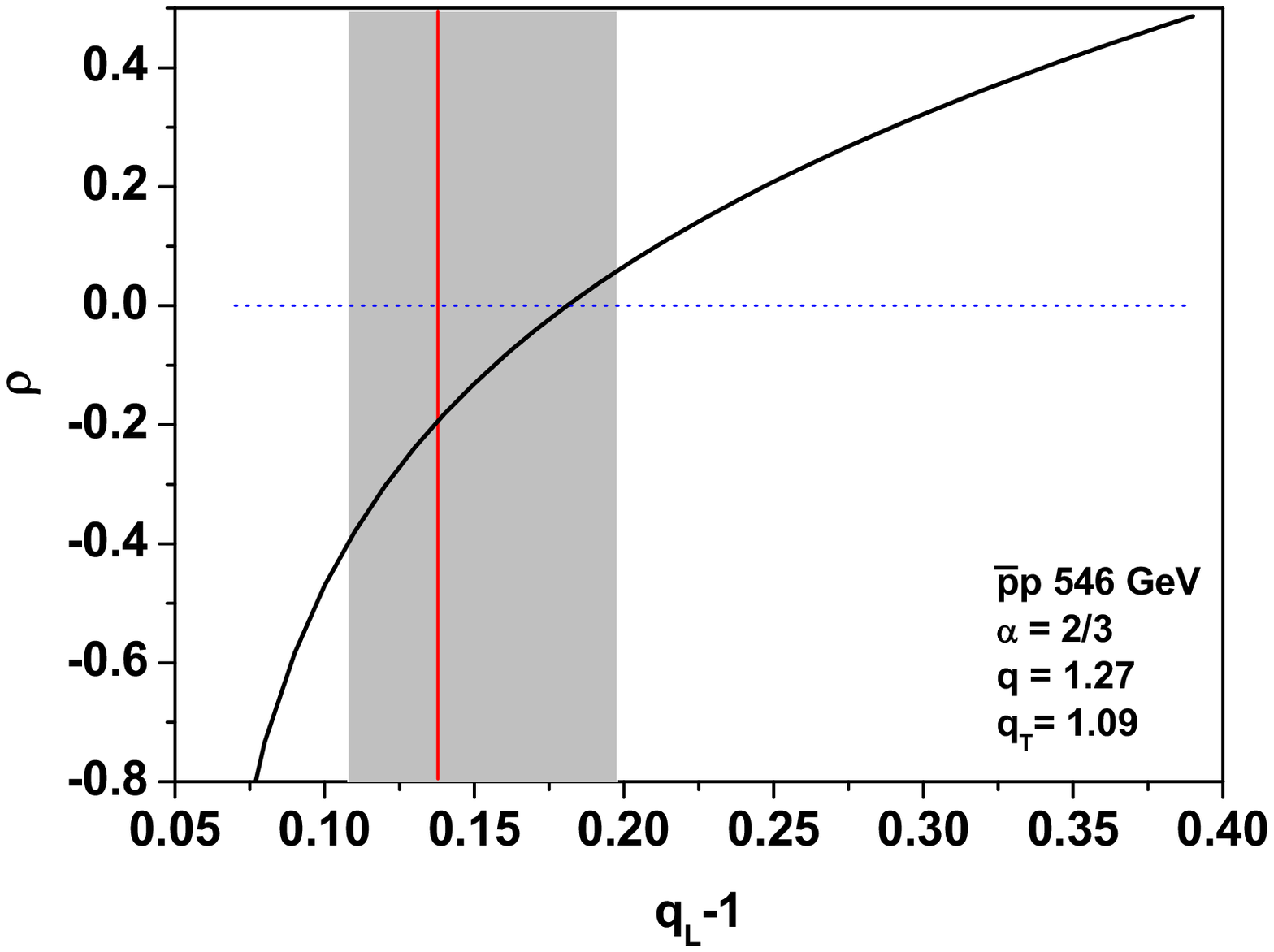}
\includegraphics [width=7.4cm]{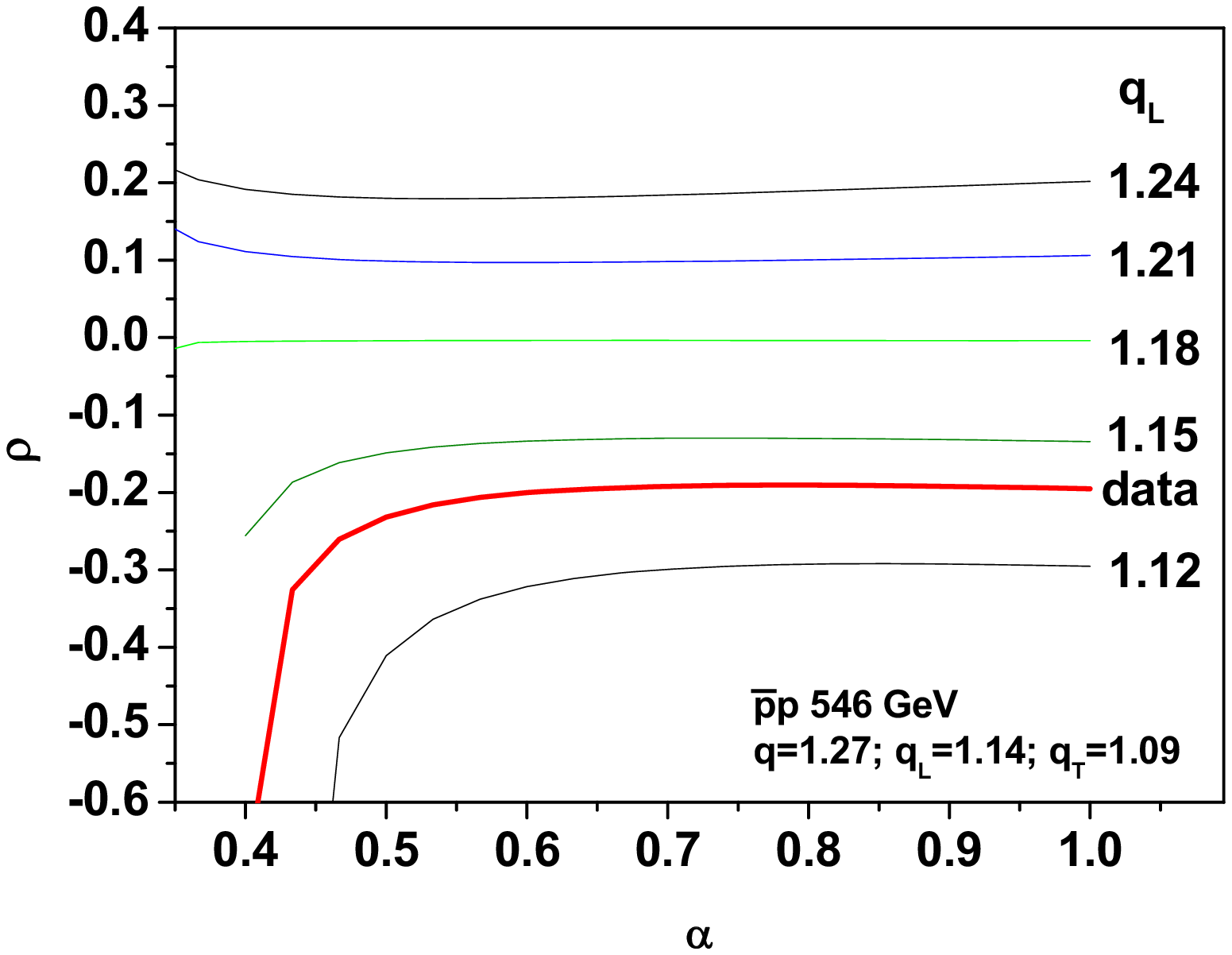}
\caption{Left panel: the example of $\rho$ obtained from Eq.
(\protect\ref{eq:rho}). The shaded area shows the extent of the
possible error, due to uncertainty in fixing $q_L$. Right panel:
the sensitivity of the correlation parameter $\rho$ to $\alpha $,
specifying the fraction of fluctuations which goes into the
transverse direction (in the case of full isotropy one has $\alpha
= 2/3$).} \label{FigB}
\end{center}
\end{figure}

In this presentation we shall not go into the details of Tsallis
statistics and its applications in the field of multiparticle
production observed in high energy experiments (see
\cite{WW,Tsallis}). Here we concentrate on the problem of the
coexistence and interconnection between fluctuations observed in
different observables and, in particular, on the possible
correlation between them which can, presumably, be measured
experimentally. In any collision one observes a number $N$ of
secondaries distributed in phase space $\{ y, p_T\}$ (with
rapidity $y$ and transverse momentum $p_T$)\footnote{Where $y =
(1/2)\ln\left[\left(E + p_L\right)/\left( E - p_L\right)\right]$,
with $E$ being energy of particle and $p_L$ its longitudinal
momentum.}. The multiplicity distributions, $P(N)$, rapidity
distributions, $dN/dy$ and distributions of transverse momenta,
$dN/dp_T$ (or, transverse masses $\mu_T = \sqrt{m^2 + p^2_T}$) are
usually measured. Among them, only $P(N)$ refers to the entire
phase space, the other are limited either to its longitudinal
($y$) or transverse ($p_T$) parts. Therefore it is natural that
nonextensivity parameters describing fluctuations present in these
observables (visible either as a broadening of $P(N)$ or
power-like tails in other distributions) are different. This is
clear from Fig. \ref{FigA} where we show that, whereas
fluctuations in full phase space depends strongly on energy, those
confined to its transverse part depend on energy weakly.
Fluctuations in the longitudinal part of phase space represent a
special case as they strongly depend on data from which they are
deduced and also on whether they are corrected for simultaneous
fluctuations in $p_T$ or not \cite{WWW}. This is shown in Fig.
\ref{FigA} by the star symbols. As shown in \cite{WWcov}, all
these fluctuations can be connected when one considers an ensemble
in which all variables characterizing an event, namely, the energy
$U$, temperature $T$ and multiplicity $N$, can fluctuate with
relative variance $\omega^2_X = Var(X)/<X>^2$, $X = U,~T,~N$. In
this case, it occurs in a natural way that a correlation
coefficient $\rho = Cov(U,T)/\sqrt{Var(U)Var(T)}$ between the two
of them (here $U$ and $T$ chosen) is necessary to fully describe
the event. Generalizing the Lindhard thermodynamic uncertainty
relations \cite{L} one obtains \cite{WWcov}:
\begin{figure}[h]
\begin{center}
\includegraphics [width=11.5cm]{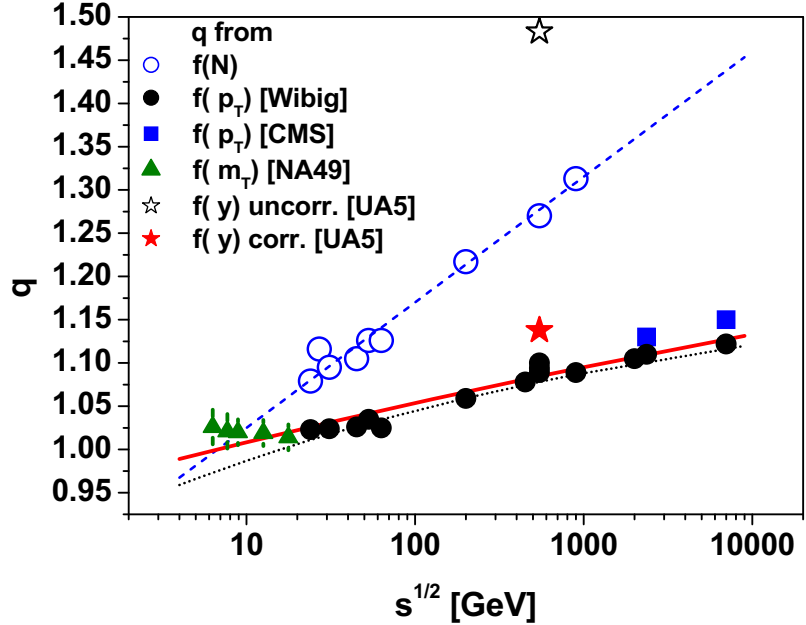}
\caption{Examples of energy dependence of the nonextensivity
parameter $q$ as obtained from different observables. Open symbols
show $q$ obtained from multiplicity distributions $P(N)$ (fitted
by $q = 1 + 1/k$ with $1/k = -0.104 + 0.029 \ln s $) \cite{PN}.
Solid symbols show $q = q_T$ obtained from a different analysis of
transverse momentum distributions, $f(p_T)$. Data points are taken
from, respectively, \cite{Wibig} for [Wibig], \cite{CMS} for
[CMS], from \cite{NA49} for [NA49] (data on $\mu_T$) and from
\cite{UA5} for [UA5]. The dotted line represents a fit obtained in
\cite{Wibig} (where $q_T = 1.25 - 0.33s^{-0.054}$) and the full
line comes from Eq. (\protect\ref{eq:all}) (for $\rho = 0$). Stars
show $q = q_L$ obtained from $dN/dy$: open star taken from
\cite{UA5} (the same as used in \cite{RWW}) shows the uncorrected
$q$ value, whereas the solid star indicates its corrected value
\cite{WWW}.} \label{FigA}
\end{center}
\end{figure}
\begin{equation}
q - 1 = \left| \omega^2_N - \frac{1}{\langle N\rangle}\right| =
\omega_U^2 + \omega_T^2 - 2\rho\omega_U\omega_T. \label{eq:qtot}
\end{equation}
What does the $\rho$ tell us? For example, a large energy $U$
(i.e., large inelasticity of reaction, $K$) can result either in a
large number of secondaries of lower energies (for which $\rho <
0$) or a smaller number of larger energies   (for which $\rho >
0$, different models give different predictions). Eq.
(\ref{eq:qtot}) tells us that, in principle, the coefficient
$\rho$ is a function of all the nonextensivity parameters
involved. Let us then specify the part of fluctuations of $T$
which go in the transverse direction by $\alpha$, and write
\begin{equation}
q_T - 1 = \alpha \omega^2_T,\qquad q_L - 1 = \omega^2_U + (1 -
\alpha) \omega^2_T.  \label{eq:qTL}
\end{equation}
From (\ref{eq:qTL}) one finds
\begin{equation}
\omega^2_T = \frac{1}{\alpha} \left(q_T - 1\right),\qquad
\omega^2_U = \left( q_L - 1\right) - (1 - \alpha)\omega^2_T =
\left( q_L - 1\right) - \frac{1 - \alpha}{\alpha} \left( q_T -
1\right).              \label{eq:omTU}
\end{equation}
Then, from (\ref{eq:omTU}) and (\ref{eq:qtot}) one gets
\begin{equation}
q - 1 = \left(q_L - 1\right) + \left(q_T - 1 \right) - 2 \rho
\omega_U \omega_T \qquad \stackrel{\rho \rightarrow
0}{\Longrightarrow} \qquad q - 1 = \left( q_L - 1 \right) + \left(
q_T - 1 \right) .\label{eq:qqTqL}
\end{equation}
Assuming that
\begin{equation}
\omega^2_U = \kappa^2 \omega^2_T ,      \label{eq:kappa}
\end{equation}
Eq. (\ref{eq:qqTqL}) reads\footnote{In \cite{WWcov} we used
$\alpha =2/3$ and $\kappa = 1$; for $\rho = 0$, the actual values
of $\alpha$ and $\kappa$ parameters are irrelevant.}
\begin{equation}
q - 1 = \left( q_L - 1\right) + \left( q_T - 1\right) -
2\rho\frac{\kappa}{\alpha}\left(q_T - 1\right) \label{eq:all}
\end{equation}
whereas from (\ref{eq:kappa}) and (\ref{eq:omTU}) one has that
\begin{equation}
\kappa = \frac{\omega_U}{\omega_T} = \sqrt{\alpha \left( \frac{q_L
- 1}{q_T - 1} + 1 \right) - 1}. \label{eq:kovera}
\end{equation}
That results in a useful correlation coefficient given in terms of
different (in principle {\it measured}) fluctuations:
\begin{equation}
\rho = \frac{ 1 - \frac{(q - 1) - \left( q_L - 1 \right)}{ q_T - 1
} }{ \frac{2}{\alpha} \sqrt{ \alpha\left( \frac{q_L - 1}{q_T - 1}
\right) - 1 }};\qquad \alpha = \frac{q_T - 1}{\omega_T^2}.
\label{eq:rho}
\end{equation}
Notice:
\begin{itemize}
\item to get the correlation coefficient $\rho$, one has to know
{\it all the fluctuations}, i.e., both in the entire phase space,
$q$, as separately in its transverse, $q_T$, and longitudinal,
$q_L$, parts;

\item out of them the best known is $q$ (no corrections needed);

\item for  $q_T$ the corrections are small and can be neglected;

\item however, for $q_L$ the corrections are large and must
  be accounted for (cf., Fig. \ref{FigA}).
\end{itemize}
The example of feasibility of deducing $\rho$ from data is
presented in Fig. \ref{FigB} for data on $\bar{p} + p$ at $546$
GeV \cite{UA5}. In this case from $P(N)$ one gets $q = 1.27$, from
distribution of $p_T$ one gets $q_T = 1.09$ whereas from the
original $q_L = 1.48$ one gets after correction $q_L = 1.14$ (cf.
Fig. \ref{FigA}).

To summarize, one can say that it is a priori possible to obtain
$\rho$ from experimental data, but - as shown here, this is
difficult. Because of large estimation errors one cannot at this
stage exclude the possibility that simply $\rho = 0$. Our
preliminary result, cautiously indicating that, perhaps, $\rho <
0$, would prefer the production of a large number of particles of
lower energies. However, statistically we cannot at this moment
make a decisive statement.

\section{Summary}

We discuss the hadronization process where all variables
characterizing an event, namely the energy $U$, temperature $T$
and multiplicity $N$ fluctuate. In this case, it follows that
correlation $\rho$ between two of them (here chosen as $U$ and
$T$), is necessary to fully describe the event. The question is:
what is its meaning and can it be estimated experimentally? The
answer to the first is that this parameter tells us how the
available energy is used: either for production of particles or,
rather, for making them more energetic. In what concerns the
second, our preliminary result (indicating that, perhaps, $\rho <
0$ ) would prefer the first scenario, however, statistically we
cannot say so at the moment.

We conclude with two remarks. First, in what concerns Eq.
(\ref{eq:qtot}), in the literature \cite{old} there is similar
relation connecting the volume, $V$, pressure, $P$ and
temperature, $T$: $\omega^2_P = \omega^2_V + \omega^2_T$
\cite{old}. Secondly, when all variables, $U$, $N$ and $T$
fluctuate, the pairs of variables, $(U,N)$ and $(U,T)$, cannot all
be independent because
\begin{equation}
Var(U) = \langle T\rangle Cov(U,N) + \langle N\rangle Cov(U,T)
\label{eq:UTN}
\end{equation}
(cf., \cite{WWcov}). This means that, in general,
\begin{equation}
\omega_U = \rho(U,N) \omega_N + \rho(U,T) \omega_T .
\label{eq:general}
\end{equation}
where $\rho(X,Y)$ denotes the corresponding correlation
coefficients between variables $X$ and $Y$ \footnote{For the
completeness of presentation one must also mention that the notion
of Tsallis statistics still remains a subject of hot debate (cf.,
for example, \cite{debate}). Nevertheless, at the moment it is
proved \cite{M} that fluctuation phenomena can be incorporated
into a traditional presentation of thermodynamics and result in
many new admissible distributions satisfying thermodynamical
consistency conditions among which is the Tsallis distribution
(\ref{eq:Tsallis}). We close this remark by noting that a recent
generalization of classical thermodynamics to a nonextensive case
\cite{recentB} clearly demonstrates its feasibility.}.\\

Acknowledgment: Partial support (GW) of the Ministry of Science
and Higher Education under contract DPN/N97/CERN/2009 is
gratefully acknowledged. We would like to warmly thank Dr Eryk
Infeld for reading this manuscript.


\begin{thebibliography}{99}

\bibitem{GGS} M. Ga\'zdzicki, M. Gorenstein, P. Seyboth, Acta Phys. Polon.
              B {\bf 42}, 307 (2011).

\bibitem{WW} G. Wilk, Z. W\l odarczyk, Eur. Phys. J. A {\bf 40}, 299
             (2009)  and Central Europ. J. Phys., DOI:10.2478/s11534-011-0111-7
             (in press, cf. also  arXiv:1110.4220v2 [hep-ph]).

\bibitem{WWq} G. Wilk, Z. W\l odarczyk, Phys. Rev. Lett. {\bf 84},
              2770 2000) and Chaos, Solitons Fractals {\bf 13}, 581
              (2001).

\bibitem{BJ} T. S. Bir\'o, A. Jakov\'ac, Phys. Rev. Lett. {\bf 94}, 132302
             (2005); T. S. Bir\'o, G. Purcel, K. \"Urm\"osy, Eur. Phys.
              J. A {\bf 40}, 325 (2009).

\bibitem{SuperS} C. Beck, E. G. D. Cohen, Physica A {\bf 322}, 267
                 (2003); F. Sattin, Eur. Phys. J. B {\bf 49}, 219
                 (2006).

\bibitem{B} T. S. Bir\'o, {\it Is there a temperature? Conceptual Challenges at
            High Energy, Acceleration and Complexity}, (Springer 2011).

\bibitem{PHENIX} A. Adare et al. (PHENIX Collaboration), Phys. Rev. D 83, 052004 (2011)

\bibitem{CMS}  V. Khachatryan et al. (CMS Collaboration), JHEP{\bf 02}, 041 (2010),
               and Phys. Rev. Lett. {\bf 105}, 022002 (2010).

\bibitem{Tsallis} C. Tsallis, Stat. Phys. {\bf 52}, 479 (1988), Eur. Phys. J. A
                 {\bf 40}, 257 (2009) and {\it Introduction to Nonextensive Statistical
                  Mechanics} (Springer, 2009).

\bibitem{WWW} G. Wilk, Z. W\l odarczyk, W. Wolak, Acta Phys. Polon. B {\bf 42},
              1277 (2011).

\bibitem{WWcov} G. Wilk, Z. W\l odarczyk, Physica\ A {\bf 390}, 3566
                (2011).

\bibitem{L} J. Lindhard, {\it 'Complementarity' between energy and
             temperature}, in  {\it The Lesson of Quantum Theory}, edited by J. de
             Boer, E. Dal,  O. Ulfbeck (North-Holland, Amsterdam, 1986);
             J. Uffink, J. van Lith, Found. Phys. {\bf 29}, 655
             (1999);

\bibitem{PN} A. K. Dash, B. M. Mohanty, J. Phys. G {\bf 37},
             025102 (2010); see also C. Geich-Gimbel,
             Int. J.Mod. Phys. A {\bf 4}, 1527 (1989).

\bibitem{Wibig} T. Wibig, J. Phys. G {\bf 37}, 115009 (2010).

\bibitem{NA49} C. Alt et al., Phys. Rev. C {\bf 77}, 034906 (2008)
               and Phys. Rev. C {\bf 77}, 024903 (2008); S. V. Afanasiev et al.,
               Phys. Rev. C {\bf 66}, 054902 (2002).

\bibitem{UA5} G. J. Alner et al. (UA5 Coll.), Z. Phys. C {\bf 33}, 1 (1986).

\bibitem{RWW} M. Rybczy\'nski, Z. W\l odarczyk, G. Wilk, Nucl. Phys. B (
              Proc. Suppl.) {\bf 122},  325 (2003); F. S. Navarra, O. V.
              Utyuzh, G. Wilk, Z. W\l odarczyk, Phys. Rev. D {\bf 67}, 114002 (2003).

\bibitem{old} Kulesh Chandra Kar, Phys. Rev. {\bf 21}, 672 (1923).

\bibitem{debate} M. Nauenberg, Phys. Rev. E {\bf 67}, 036114 (2003);
                 Phys. Rev. E {\bf 69}, 038102 (2004); C. Tsallis,
                 Phys. Rev. E {\bf 69}, 038101 (2004);
                 R. Balian, M. Nauenberg, Europhys. News 37, 9 (2006);
                 R. Luzzi et al. Europhys. News 37, 11 (2006).

\bibitem{M} O. J .E. Maroney, Phys. Rev. E {\bf 80}, 061141
            (2009).

\bibitem{recentB} T. S. Bir\'o, K. \"Urm\"ossy, Z. Schram, J. Phys. G {\bf 37},
                  094027 (2010); T. S. Bir\'o, P. V\'an, Phys. Rev. E {\bf 83},
                  061147 (2011);  T. S. Bir\'o,  Z. Schram, EPJ Web of Conferences
                  {\bf 13}, 05004 (2011).

\end{thebibliography}
\end{document}